\documentclass[aps,pra,twocolumn,showpacs,showkeys]{revtex4}
\usepackage{graphicx}
\usepackage{graphics}
\usepackage{color}  
\usepackage{subfigure}
\usepackage{tikz}
\usetikzlibrary{shapes}
\usetikzlibrary{fadings}
\usetikzlibrary{shadows}

\begin{document}
\title{Creation and characterization of vortex clusters in atomic Bose-Einstein condensates}
\author{Angela C. White}
\email{ang.c.white@gmail.com}
\author{Carlo F.~Barenghi}
\author{Nick P.~Proukakis}
\affiliation{School of Mathematics and Statistics, Newcastle University, Newcastle upon Tyne, NE1 7RU, England, UK.}
%
\begin{abstract}
We show that a moving obstacle, in the form of an elongated paddle, can create vortices that are dispersed, or induce clusters of like-signed vortices in 2D Bose-Einstein condensates. 
We propose new statistical measures of clustering based on Ripley's K-function which are suitable to the small size and small number of vortices in atomic condensates, which lack the huge number of length scales excited in larger classical and quantum turbulent fluid systems.  The evolution and decay of clustering is analyzed using these measures.  Experimentally it should prove possible to create such an obstacle by a laser beam and a moving optical mask.  The theoretical techniques we present are accessible to experimentalists and extend the current methods available to induce 2D quantum turbulence in Bose-Einstein condensates. 
\end{abstract}


\maketitle

\section{Introduction}
Turbulent superfluid helium has been shown to exhibit features typical of ordinary, classical turbulence, on scales larger than the average inter-vortex spacing, such as the same Kolmogorov energy spectrum in three dimensional (3D) turbulence \cite{Nore1997,Maurer1998,Stalp1999,Araki2002,Kobayashi2005,Salort2010PF,Baggaley2011PRE,Baggaley2011PRB} as well as differences, such as non-Gaussian velocity statistics \cite{Paoletti2008, White2010, Adachi2011,Baggaley2011PRE}.  Two dimensional (2D) turbulence is very different from 3D turbulence.  In 3D turbulence, on scales larger than the average inter-vortex spacing,  there is a flow of energy from large scales to small scales through the Richardson cascade, whereby large eddies are broken up into smaller and smaller eddies giving a Kolmogorov energy spectrum with $k^{-5/3}$ scaling. In 2D turbulence there is an inverse energy cascade, where energy flows from the scale of energy injection (small scales), to larger scales as like-charged vortices cluster \cite{Kraichnan1967}.  This phenomenon is established in classical fluids and is thought to be the mechanism responsible for Jupiter's great red spot  \cite{Marcus1988,Sommeria1988}.  
The concept of an inverse energy cascade was introduced by Onsager \cite{Onsager1949} for a point-vortex gas consisting of many many vortices, where it was found clusters of like-signed point vortices have a negative temperature (further detail can be found in subsequent analyses \cite{Montgomery1972,Montgomery1974}).  Beyond the seminal work of Onsager, there is no fundamental theory that exists for small systems of vortices.  In Bose-Einstein condensates, while there have been a number of theoretical investigations \cite{Parker2005,Horng2009, Numasato2010JLTP,Numasato2010} and some experimental work \cite{Neely2012}, the question of whether an inverse cascade is a feature of a system of turbulent vortices in quasi-two dimensional systems remains open; this is firstly because these condensates are relatively new and secondly (and more importantly) because, due to their relatively small size, they lack the large range of length scales typical of other 2D turbulent flows (such as planetary atmospheres).  

On the other hand, ultra-cold Bose condensed atomic gases are ideal to investigate the dynamics of few-vortex systems as well as turbulent dynamics of many vortices.  The dimensionality of such condensates can be easily controlled, allowing direct study of vortices in two and three dimensions.  Advanced experimental methods for the imaging and detection of vortices in Bose--Einstein condensates (BECs) have been developed \cite{Freilich2010} and new theoretical proposals have shown that vortices can be easily manipulated in BECs \cite{Aioi2011}. There are many techniques that can be applied to nucleate vortices in BECs, such as: engineering the condensate phase profile \cite{Ketterle2002,Shibayama2011}; stirring the condensate with a blue or red detuned laser (for experiments see \cite{Inouye2001,Neely2010} and theory \cite{Fujimoto2010,Fujimoto2011,Aioi2011}); mixing and merging condensates of well defined phases \cite{Scherer2007,Carretero2008,Ruben2008}; moving a condensate past a defect \cite{Neely2010}; rotating the trapping potential or thermal cloud \cite{Madison2000,Raman2001,Ketterle2001,Hodby2001,Cornell2003,Fetter2009};  and cooling the condensate with a rapid quench through the phase transition (Kibble-Zurek mechanism) \cite{Kibble1976,Zurek1985,Weiler2008}.  Finally, vortices can be nucleated from dynamical instabilities, such as through the decay of the snake instability of a soliton \cite{Anderson2001, Dutton2001}, the bending wave instability of a vortex ring \cite{Horng2006,Horng2008} or surface mode excitations of the condensate \cite{Henn2009,Henn2010}. In the first experimental demonstration of 3D quantum turbulence in a BEC, tangled vortices were created by shaking the condensate with an oscillatory trapping potential \cite{Henn2009,Henn2010, Seman2011LPL, Shiozaki2011LPL}. 

In this paper we add to the numerous existing techniques applied to induce vortices in BECs, and apply a moving object with an elliptical paddle shape to create vortices. We demonstrate that the trajectory of the optical paddle through the condensate can be controlled to create vortices that are both well distributed or clustered into groups with like-winding.  To determine if vortices are indeed clustered, and if this clustering increases or decreases significantly with time, we develop tools which are more suitable to the small size and the relative small number of vortices which can be generated in ultracold atom BECs.  Drawing on the wealth of available statistical pattern recognition methods,  we analyze our data using Besag's function \cite{Besag1977}, a modification of Ripley's K-function \cite{Ripley1976,Ripley1977}, which has been extensively applied across a variety of scientific fields to measure clustering and clumping of discrete objects e.g. \cite{Pelissier1999,Marcon2003,Lancaster2004,Jones2010,Jafari2010,Kiskowski2009,Baggaley2011}.  Motivated by Besag's function, we develop some new measures of independent clustering when the system is comprised of two unique types of discrete objects.  In our case, these are vortices with `$+$' or `$-$' winding in a BEC.  These techniques can distinguish between the cases of mixed clusters of vortices and independent clusters of vortices of like-winding in the condensate.  After reviewing our theoretical model, we analyze vortex structures obtained by different forms of stirring by applying both Besag's function and nearest neighbor techniques. 
Finally, we summarize our main conclusions, that the resulting vortex clustering is strongly dependent on the trajectory of the moving paddle. We find no compelling evidence of an inverse cascade in these small systems, in the sense that we don't see an increase in clustering over time. 
\section{Numerical Model}
We simulate the dynamics of a trapped two-dimensional Bose-Einstein condensate stirred by an optical paddle by integrating the following 2D dimensionless time-dependent Gross-Pitaevskii equation (GPE):
  \begin{equation}\label{gpe}
i \frac{\partial \psi}{\partial t} = -\frac{1}{2}\nabla^{2}\psi + \frac{{r}^{2}}{2}\psi+V_{P}\psi + \kappa_{2d} |\psi|^{2}\psi\,,
 \end{equation}
where the interaction strength, $\kappa_{2d}= 2\sqrt{2\pi}aN / a_{z} $, is written in terms of the scattering length, $a$, and the total number of condensate atoms, N.  $\psi$ is scaled as $\psi=a_{r}\Psi/\sqrt{N}$, where $\Psi$ is the condensate wavefunction and is normalized to unity i.e. $\int\text{d\bf{x}}|\psi|^{2}=1$.  Lengths and time are scaled as $\tilde{r}/a_{r}=r$ and $\tilde{t}\omega_{r}=t$, where $\tilde{r}$ and $\tilde{t}$ are dimensional with units of meters and seconds respectively.   $a_{z}= \sqrt{\hbar/(m\omega_{z})}$ and $a_{r}= \sqrt{\hbar/(m\omega_{r})}$ are the axial and radial harmonic oscillator lengths, determined by the axial and radial trapping frequencies, $\omega_{z}$ and $\omega_{r}$ respectively.

The potential, $V_{P}$, describes a far-off-resonance blue-detuned laser beam shaped into a paddle that follows a rotating and circular stirring trajectory given by \begin{eqnarray}\label{paddle}\nonumber
V_{P} = V_{0}\exp\left[
-{\eta}^{2}\frac{\left(\tilde{x}\cos(\omega t)-\tilde{y}\sin(\omega t)\right)^{2}}{d^{2}} \right. \\ \left. - \frac{\left(\tilde{y}\cos(\omega t)+\tilde{x}\sin(\omega t)\right)^{2}}{d^{2}}\right]\,,
\end{eqnarray}
where $\tilde{x} = x-v\sin(t)$ and $\tilde{y}=y-v\cos(t)$. $V_{0}$ gives the peak strength of the potential and is selected to be $V_{0}\sim 2.6 \mu$, where $\mu$ is the chemical potential of the BEC. $\eta$ determines the paddle elongation and $d$ the width.  Experimentally, the paddle can be shaped by shining a far-off-resonance blue detuned laser through a mask as in \cite{Dennis2010}.  For a paddle rotating with a frequency, $\omega$, at the center of the condensate, $\tilde{x} = x$ and $\tilde{y} = y$. For a paddle moving at a constant radius from the center of the condensate without rotating, we take $\omega =1$. In all simulations the paddle is initially linearly ramped up to its maximum stirring frequency, $\omega$, after which the condensate is stirred at constant $\omega$ until $t_{S}=20$. The paddle is then ramped off linearly over $t=5$ by making the replacement 
\begin{equation}
V_{P} \rightarrow \left(1-\frac{(t-t_{S})}{5}\right)V_{P},
\end{equation}
in Eq.~(\ref{paddle}).  For paddle sizes and stirring frequencies which result in vortex formation, the condensate dynamics are evolved for a further time of $t=45$.

Numerically, Eq.~(\ref{gpe}) is solved pseudo-spectrally with periodic boundary conditions and integrated in time by applying an adaptive 4th-5th order Runge-Kutta method with the help of xmds \cite{xmds}. The initial state for our simulations is obtained by a short propagation of the 2D GPE  in imaginary time (by making the replacement $\tau = -it$ in Eq.~(\ref{gpe})), and applying a stationary paddle potential. The condensate is parameterized by the nonlinear interaction strength $\kappa_{2d} = 10399$.  
For example, by choosing experimentally relevant axial and radial trapping frequencies of  $\omega_{z}=2\pi \times50$ Hz and $\omega_{r}=2 \pi \times 5$ Hz, this describes a condensate of $2.2\times10^{6}$ $^{23}$Na ($6\times10^{5}$ $^{87}$Rb) atoms  with scattering length $a = 2.75$nm ($a=5.29$nm). Our simulations are run on a grid of spatial extent $-20$ to $20$ with gridsize $N_{g}=512$ (see appendix). 

\section{Vortex Generation}

This paper looks at three stirring motions of the paddle, $V_{\text{I}}$, $V_{\text{II}}$ and $V_{\text{III}}$, each of which are specific realizations of Eq.~(\ref{paddle}), corresponding to a rotating paddle, a paddle moving on a deferent from the condensate center and a paddle moving along a deferent while rotating, respectively (see figure 1, top row). The stirring motions of the paddle studied generate contrasting vortex configurations, as discussed below. For each case,  we track the total number of vortices nucleated, as shown in figure 3a. In all cases, at later times, vortices are lost to the edge of the condensate. The condensate edge is selected by identifying where the density falls to less than $30\%$ of the maximum condensate density at that time. The resulting profile is then smoothed to give the condensate edge.  When two vortices are closer than a critical separation distance and are of opposite winding, vortex pair annihilation occurs, a mechanism which also reduces the total number of vortices in the condensate. 

For all simulations  the total angular momentum of the condensate is also tracked, given by 
\begin{equation}\label{angmom}
\langle L_{z} \rangle = -i\int\text{d\bf{x}}\psi^{*}\left(x\frac{\partial}{\partial y}-y\frac{\partial}{\partial x}\right)\psi\,.
\end{equation}

  \newpage
{
 \onecolumngrid
 
{\begin{figure}[t!]\label{paddlecollated.jpg}
\includegraphics[width = \linewidth]{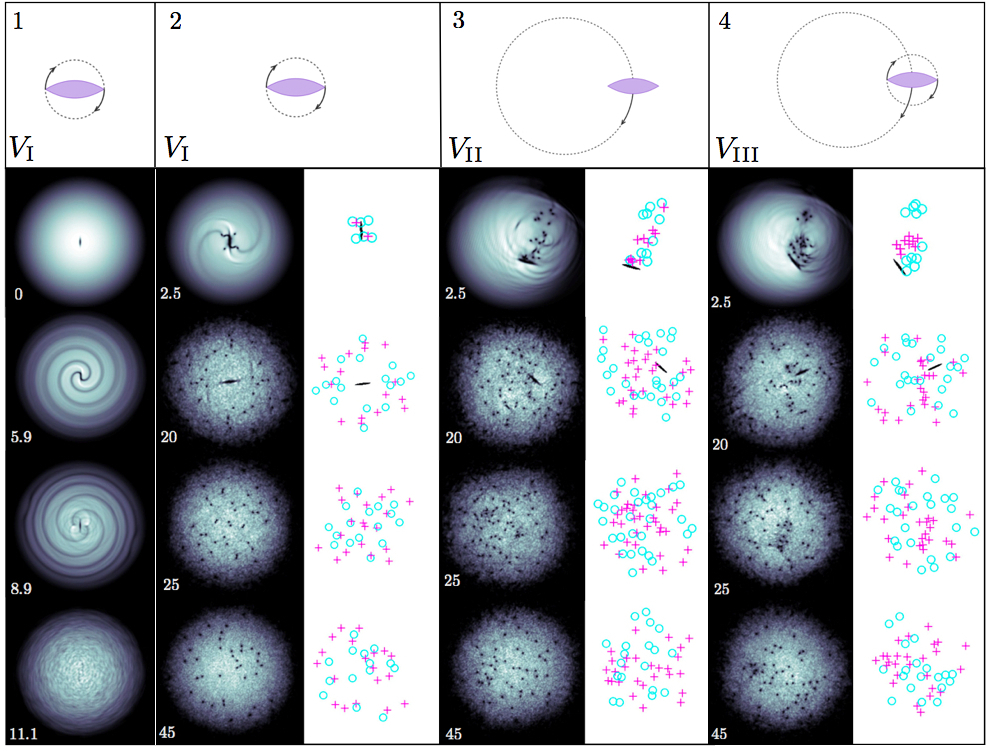}
\caption{(Color online) Column 1: Density slices for paddle $V_{\text{I}}$ with parameters: $d=0.5$, $\omega =6$ (dimensionless units); Columns 2--4: Density and vortex position images for paddles $V_{\text{I}}$, $V_{\text{II}}$ and $V_{\text{III}}$.  Parameters: $V_{\text{I}}$ (column 2): $d=1$, $\omega=4$; $V_{\text{II}}$ (column 3): $d=1$, $\omega=1$, $v=4$; $V_{\text{III}}$ (column 4):  $d=1$, $\omega=2$, $v=4$.  All paddles have $V_{0} =150$, $\eta=8$ (dimensionless units). Positive vortices are identified by (pink) $\color{magenta}{+}$ signs and negative vortices by (blue) $\color{cyan}{\circ}$ signs. Time is indicated in white at the bottom left hand side of each plot. }
\end{figure}}

 \twocolumngrid}
\noindent   
\subsubsection*{Case I: Paddle rotating at the condensate center}
Firstly we look at a paddle rotating about its center with frequency $\omega$ at the center of the condensate. This is modeled by evolving Eq.~(\ref{gpe}) with $V_{\text{I}}=V_{P}(\tilde{x}\rightarrow x, \tilde{y}\rightarrow y)$.  The smallest paddle size we consider is $d=0.5$, rotating at a frequency $\omega = 6$.  The rotating motion of the paddle produces circular spiral sound waves, which at late times interfere with each other giving a wave interference pattern (see the first column of figure 1). Paddles with a larger width ($d=1$), rotating at frequencies of $\omega = 4, 6$ and $8$, nucleate vortices in addition to creating spiral sound waves. The density profile for a paddle rotating at frequency $\omega = 4$ is show in the second column of figure 1 (see supplementary movie 1 for a comparison of paddles rotating at frequencies $\omega = 4$ and $\omega = 8$).  A greater rate of rotation increases the number of vortices initially nucleated, as expected. In all cases, vortices are initially nucleated from the ends of the paddle with winding opposite to the direction of rotation of the paddle, as depicted in figure 2. At subsequent times, when the local superfluid velocity surpasses the critical velocity for vortex nucleation \cite{Frisch92}, vortices of both signs are nucleated from both the center and ends of the paddle. A centered rotating paddle imparts a small amount of angular momentum to the condensate.  The angular momentum imparted is proportional to the frequency of the rotating paddle (see figure 3b). Note that after paddles of frequency $\omega = 6$ and $8$ have been ramped off, the condensate angular momentum saturates to approximately the same value.  

\begin{figure}
\includegraphics[width=\linewidth]{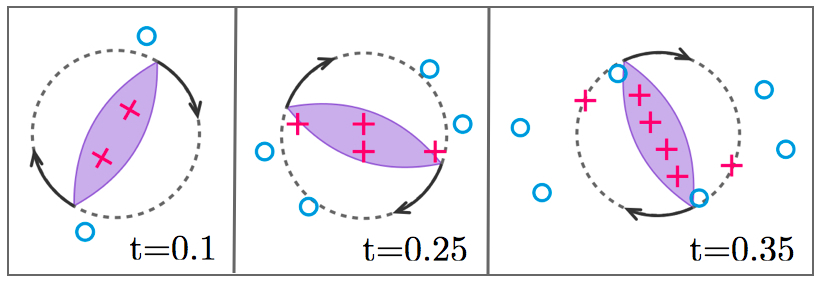}
\caption{(Color online)Schematic diagram of vortices nucleated from a paddle rotating at the condensate center, $V_{\text{I}}$ as the paddle is ramped up to $\omega=8$. Positive vortices are identified by (pink) $\color{magenta}{+}$ signs and negative vortices by (blue) $\color{cyan}{\circ}$ signs, the paddle profile is shown in purple.}
\end{figure}

{\begin{figure}\label{VortexCount}
\subfigure[Total Vortex Number]{\includegraphics[width=0.8\linewidth]{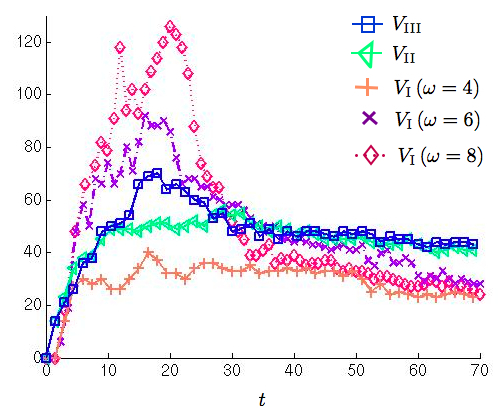}}
\subfigure[Angular Momentum]{\includegraphics[width = \linewidth]{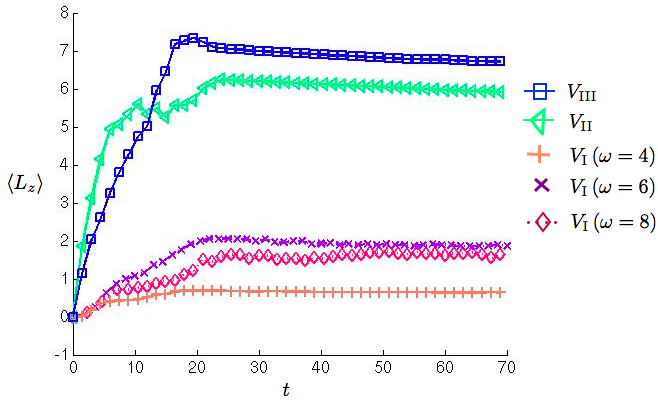}}
\caption{(Color online) Variation of a) total vortex number and d) condensate angular momentum with time.  Vortices induced by a rotating paddle, $V_{\text{I}}$, with $d=1$ and $v=4, 6$  and $8$ are represented by orange pluses, purple crosses and pink diamonds respectively. Data corresponding to the paddles with $v=4$ and $d=1$, moving with trajectories $V_{\text{II}}$ ($\omega = 1$) and $V_{\text{III}}$ ($\omega = 2$),  are given by green triangles and blue squares respectively.}
\end{figure}}

\subsubsection*{Case II: Paddle stirring at constant radius}
The second stirring motion of the laser paddle we simulate is a paddle stirring the condensate at a constant radius from the condensate center, or traveling along a deferent, $V_{\text{II}}=V_{P}(\omega=1)$ (see figure 1, column 3, top row, for a schematic of the paddle motion). This paddle trajectory creates vortices which tend to group initially in like-signed clusters as shown by the progressive time samples of the condensate density profiles in figure 1, column 3 (see also supplementary movie 2). The condensate gains a large amount of angular momentum (refer to figure 3b), and consequently at final times there is a significant imbalance in the total number of vortices with positive and negative winding. 

\subsubsection*{Case III: Paddle rotating and stirring at constant radius}
The final stirring trajectory of the laser paddle we simulate, $V_{\text{III}}=V_{P}$, is a combination of the two previous motions, with the paddle moved at a constant radius from the condensate center while rotating at a small frequency. This motion can also be described as a paddle rotating and moving along the deferent (see figure 1, column 4, top row, for a schematic of the paddle motion). The effect of adding the rotational motion of the paddle to its trajectory mixes the clusters produced, resulting in smaller groups consisting of $3$ and $4$ vortices (compare the vortex distributions in figure 1 columns 3 and 4 and supplementary movie 2).  While initially (for times $t<12$) the angular momentum transfered to the condensate is  reduced in comparison to that of paddle $V_{\text{II}}$ (see figure 3b), this stirring trajectory results in the greatest transfer of angular momentum to the condensate at later times.

\section{Clustering Analysis}

\subsection{Ripley's K-function}
We analyze the clustering of vortices formed by stirring a 2D condensate with a paddle by applying Ripley's K-function, a statistical pattern analysis method used as a measure of spatial clustering.  In the context of clustering of like-signed vortices, Ripley's K-function is dependent on the total number of liked signed vortices, $N$, within the total condensate area, $A$, and can be expressed as  
\begin{equation}
K(r) = \frac{A}{N^{2}}\sum_{i=1}^{N}\sum_{j=1}^{N}f_{ij}(r)\,, 
\end{equation}
where $f_{ij}(r) = 1$ for a vortex, $j$, within a distance $r$ of the reference vortex, $i$, with like-winding. Otherwise, $f_{ij}(r) = 0$ if  $i=j$, or if the distance between vortex $i$ and $j$ is greater than $r$. That is
\begin{equation}
f_{ij} = \left[
\begin{array}{ll}
1 & \forall \, r_{ij}<r, \, i \neq j \\
0 & \forall \, r_{ij}>r \,\,\text{or}\,\, i = j
\end{array}\right.
\end{equation}
here $r_{ij}$ is the distance from a reference vortex $i$ to the comparison vortex $j$ with like-winding. This is depicted in figure 4. 

\begin{figure}
\includegraphics[height = 0.3\linewidth]{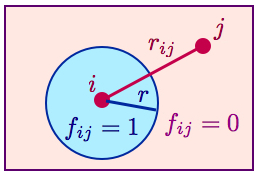}
\caption{The function $f_{ij}(r)$ in Ripley's K-function.}
\end{figure}

\begin{figure}[h!] \label{CompensatedBesag}
{\includegraphics[width=1\linewidth]{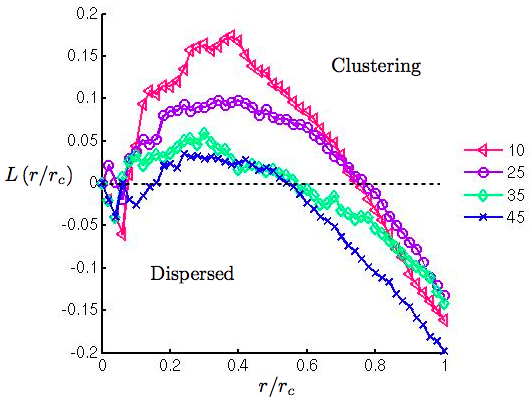}}
\caption{Besag's function $L(r/r_{c})$ (see Eq.~9)  for vortices of negative winding at varying times (see legend) for a condensate with vortices nucleated by a paddle with trajectory $V_{\text{II}}$. Parameters: $d=1$, $\omega=1$ and $v=4$.}
\end{figure}

Ripley's K-function looks at the number of like-signed vortices within a radius, $r$, from the position of an arbitrarily chosen vortex, $i$, at its center (see figure 4 and 5). If the number of  vortices with like-winding per unit area within this radius, $r$, is greater than the overall number of like-signed vortices per unit area for the whole condensate, then the vortices are said to be clustered. Clustering results in $K(r)$ increasing faster than if vortices of either sign are distributed in a spatially random manner, that is, if they follow a Poisson distribution. Ripley's-K function for a poisson-distributed data set takes the form $K(r)=\pi r^{2}$. For a linear scaling of Poisson-distributed data, it is useful to normalize Ripley's K-function to $H(r)=\sqrt{K(r)/\pi}$.  Ripley's L-function, also commonly known as Besag's function, is obtained from further normalization of Ripley's K-function: 
\begin{equation}
L(r) = \sqrt{K(r)/\pi}-r.
\end{equation}
As the condensate area, $A$, does not necessarily remain constant over all times, we scale $r$ by the characteristic condensate radius $r_{c}=\sqrt{A/\pi}$ for that time and in our subsequent analysis evaluate
\begin{equation}
L(r/r_{c}) =\sqrt{\frac{A}{\pi(Nr_{c})^{2}}\sum_{i=1}^{N}\sum_{j=1}^{N}f_{ij}(r/r_{c})}-\frac{r}{r_{c}},
\end{equation}
which simplifies to:
\begin{equation}
L(r/r_{c}) =\sqrt{\frac{1}{N^{2}}\sum_{i=1}^{N}\sum_{j=1}^{N}f_{ij}(r/r_{c})}-\frac{r}{r_{c}}.
\end{equation}
Besag's function is zero for like-signed vortices which are randomly distributed, takes positive values for vortices clustered over that spatial scale, and is negative if the vortex distribution is dispersed.  That is:
\begin{equation}
L(r/r_{c}) = \left[
\begin{array}{rl}
1 & \text{Clustered} \\
0 & \text{Random} \\
-1 & \text{Dispersed}
\end{array}\right.
\end{equation}
The radius around a centered vortex containing, on average, the most like-signed vortices per area, is called the radius of maximal aggregation, and is given by the value of $r$ which maximizes $L(r)$ \cite{Kiskowski2009}.  

{\begin{figure}[t!]
\includegraphics[height = 3cm]{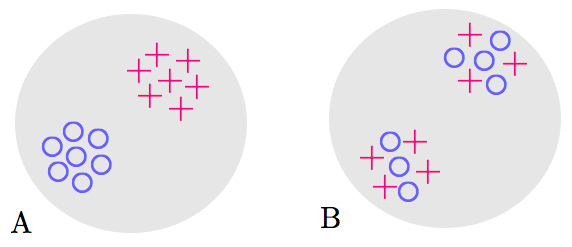}
\caption{(Color Online) Schematic diagrams of independent (A) and co-clustering (B)  in systems with two distinct types of objects, represented by pink pluses and blue circles respectively.}
\end{figure}\label{schematic}}

For a paddle rotating at a constant radius from the condensate center, evaluating $L(r/r_c)$ (see equation 9) for positive vortices, as seen in figure 5, shows that the clustering of vortices decreases with time. Although the vortices are clustered, the amount of clustering is not constant or increasing in time, giving no evidence of an inverse cascade for this system. 
{\begin{figure}[h!]\label{nearestneighbor}
\subfigure{\includegraphics[width=\linewidth]{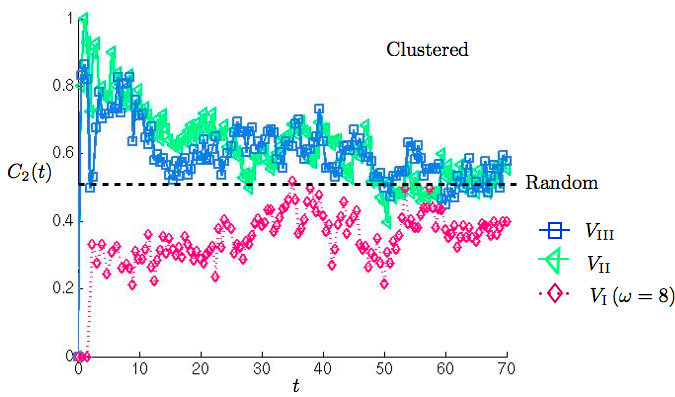}}
\subfigure{\includegraphics[width=\linewidth]{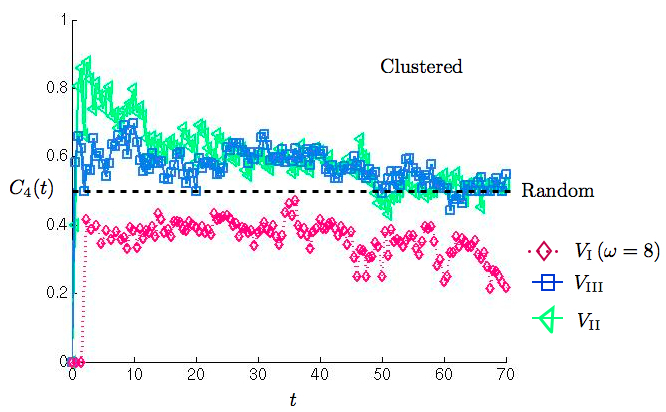}}
\caption{(Color online) Comparison of evolution second nearest neighbor data, $C_{2}(t)$, (top figure) and fourth nearest neighbor data, $C_{4}(t)$, (bottom figure) for paddle trajectories $V_{\text{II}}$, $V_{\text{III}}$ and $V_{\text{I}}$ with $v=8$ represented by green triangles, blue squares and pink diamonds respectively.  Time is measured from the beginning of stirring the condensate.}
\end{figure}}

\begin{figure}[h!]
\subfigure{\includegraphics[width=\linewidth]{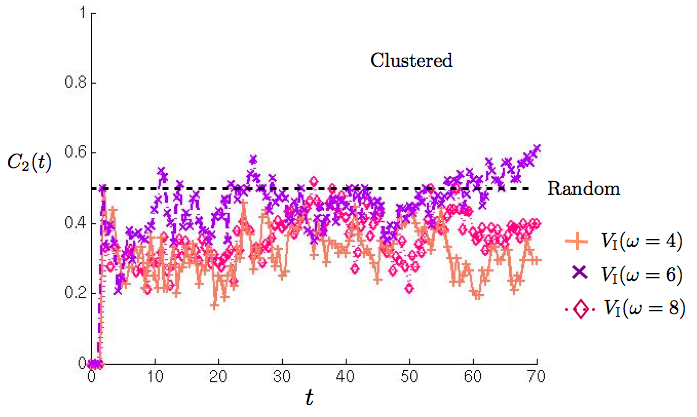}}
\subfigure{\includegraphics[width=\linewidth]{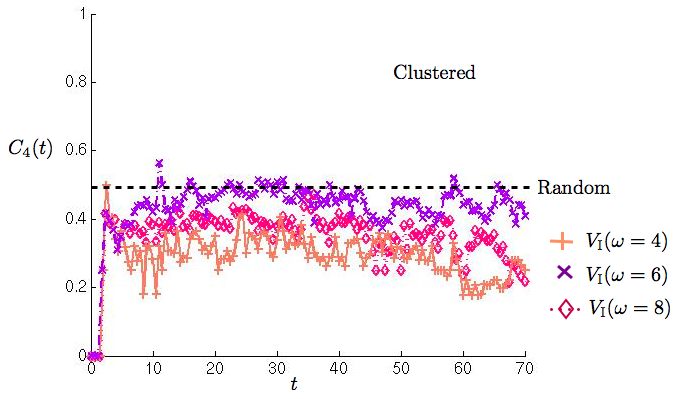}}
\caption{(Color online) Comparison of evolution second nearest neighbor data, $C_{2}(t)$ (top figure), and fourth nearest neighbor data, $C_{4}(t)$ (bottom figure), for vortices created by rotating a paddle at the condensate center ($V_{\text{I}}$) with $v=4$ (orange $+$), $v=6$ (purple crosses) and $v=8$ (pink diamonds).  Time is measured from the beginning of stirring the condensate.}
\end{figure}

\subsection{Measures of independent clustering of like-signed vortices}

While Besag's function gives a measure of the clustering of vortices with the same winding, it does not discriminate between cases where like and opposite signed vortices are clustered in the same spatial region and cases where like-signed vortices are clustered in spatially independent regions (refer to figure 6 for a schematic illustration).  It is necessary to make this distinction when looking for a measure of an inverse cascade process, as the clustering of like signed vortices is not expected to occur in the same spatial region as  clustered vortices of the opposite sign. To address this issue we define a new measure of clustering which uses the sign of nearest neighboring vortices to determine if clustering occurs in regions that are spatially independent. 

We express this measure of clustering based on looking at the sign of all $j$ neighboring vortices up to the $B$th nearest neighbor of an arbitrary reference vortex $i$ as
\begin{equation}
C_{B}(t) = \frac{1}{N}\sum_{i=1}^{N}\sum_{j=1}^{B}\frac{c_{ij}(t)}{B}\,.
\end{equation}
Here $c_{ij} = 1$ if the vortex $i$ and its $j^{th}$ nearest neighbor are of the same sign and $c_{ij} =0$ if vortex $i$ and its $j^{th}$ nearest neighbor are of opposite sign. If vortex $i$ is separated by distance greater than $R_c=r_{c}/3$ to its $j^{th}$ nearest neighbor, then $c_{ij} = 0$. $B$ is the maximum nearest neighbor to the reference vortex $i$.  It is necessary that the value of $R_{c}$ chosen is greater than the average inter-vortex separation distance and on order of the largest cluster size. Vortices closer than $R_{c}$ to the condensate edge will bias the calculation of $C_{B}(t)$ as they have an area less than $\pi R_{c}^2$ surrounding them, which their nearest neighboring vortices could inhabit.  To correct for these edge effects we omit these vortices which are less than $R_{c}$ from the condensate edge from the set of reference vortices, but still include them in the set of comparison vortices for vortices a distance greater than $R_{c}$ from the condensate edge.  Systems for which $C_{B}=0.5$ are randomly distributed and when $C_{B}$ takes values greater than $0.5$ the objects are clustered.  We note that these measures can be applied generically to investigate cases of  co and independent clustering of two discrete objects and could be simply extended to look at co and independent clustering of many discrete objects. 

A comparison of the evolution of $C_{B}(t)$ for the simulation runs described previously is shown in figures 7 and 8. From the nearest neighbor analysis we learn:
\begin{itemize} 
\item{From figure 7, we can see that a paddle moving at a constant radius from the condensate center ($V_{\text{II}}$) creates vortices that are initially very clustered, indicated by $C_{2}$ and $C_{4}$ taking values very near one. After the paddle is turned off the vortices remain clustered, with $C_{2}$ and $C_{4}$ not decreasing below $0.5$ until $t\approx 50$. At long times ($t > 50$) vortices become randomly distributed.}
\item{A paddle moving with trajectory $V_{\text{III}}$ creates clusters that are initially smaller in size than purely moving the paddle at constant radius from the condensate center ($V_{\text{II}}$). After the paddle is turned off, the vortex distribution closely follows that of run $V_{\text{II}}$ with vortices remaining clustered until long times ($t > 50$) when they become randomly distributed.  }
\item{When the paddle is only rotated at the condensate center, $V_{\text{I}}$, vortices never become clustered (refer to figure 8).}
\end{itemize}
 
 For all cases, regardless of how the vortices are initially nucleated, evaluating $C_{B}(t)$ gives no evidence of a tendency of increasing clustering of like-signed vortices over a scale of $R_{c}$ in a turbulent 2D BEC after the laser paddle has been ramped off.  Our observations imply that clusters of like-signed vortices exist due to the way in which they were induced in the condensate but do not naturally tend to cluster.  
 
\section{Conclusion}
In this paper we have covered two main objectives:
\begin{itemize}
\item{We have extended the available methods for creating vortices in 2D atomic Bose-Einstein condensates, demonstrating that a paddle can be used to stir a condensate in two quite different ways, creating long-lived vortex clusters or more randomly distributed vortices that are turbulent in two dimensions.}
\item{ A new statistical measure of clustering based on analyzing nearest neighbor vortices was defined, motivated by a well known statistical spatial point pattern analysis technique, Besag's function. These measures have been applied to analyze how vortices are distributed in 2D condensates.}
\end{itemize}

We find that a paddle moved through the condensate at a constant radius from the center creates vortices of both positive and negative winding in clusters. When the paddle is rotated at the condensate center, vortices created are initially clustered co-dependently in the same local spatial regions and later disperse throughout the condensate. For a combination of both moving a paddle at a constant radius through the condensate while simultaneously rotating the paddle, the vortices induced are less clustered than if the paddle is only moved at a constant radius from the condensate center.  The later method can be applied to create long-lived vortex clusters in BECs. The clusters are considered long lived in terms of two relevant sets of timescales; the constraining timescale determined by realistic experimental lifetimes of BECs and the timescales determined by the system size and properties.  The relevant timescales intrinsic to  the system size and length-scales are determined by the average separation distance between vortices ($l_{sep}$) and is given by $\tau_{sep} = l_{sep}^{2}/\gamma$ where $\gamma$ is the phase winding of a vortex and the largest turnover time $\tau = \pi r_{c}^{2}/2\pi \approx 35$, is determined by the condensate size. As the largest turnover time is smaller than the longevity of clusters,  which persist for $\sim 50$ (see $C(4)$ in Figure 7) we describe the clusters to be `long-lived'.  For our choice of experimental parameters, cluster lifetime is $\sim 1$s, which is also long-lived in comparison to the typical experimental lifetime of BECs, ($\sim10$s). 

The extent of clustering was quantitatively measured by evaluating two statistical measures of clustering; applying a modified Ripley's function and a technique based on comparing the sign of nearest neighboring vortices.  We did not observe an increase in clustering over time. This was despite evolution times longer than the largest turnover time  $\tau = \pi r_{c}^{2}/2\pi\approx 35$, determined by the condensate size.  Our system contains too few vortices to determine if the relevant physical process for 2D turbulent systems in atomic Bose-gases is an inverse cascade of incompressible kinetic energy from small to large scales manifesting in a clustering of like-signed vortices.  In particular, it would be difficult to apply traditional methods used for large systems (planetary atmospheres, superfluid helium), based on Fourier transforming the velocity field and analyzing the spectra of energy and enstrophy over many decades in wavenumber space.  Our statistical analysis based on Ripley's K-function and on nearest neighbor methods provides a way to quantify an increase or decrease in the degree of vortex clustering. As these methods are constructed from a knowledge of the position and winding of vortices in the system they are readily accessible experimentally. Information on vortex location in condensates can be obtained experimentally through standard absorption imaging techniques, e.g. \cite{Madison2000,Raman2001,Freilich2010}, and winding of vortices is found by analysis techniques giving phase information, such as condensate interferometry, e.g. \cite{Chevy2001,Inouye2001,Had2006}.  

{\acknowledgements A.~White thanks C.~J.~Foster for the vortex detection algorithm based on the plaquette technique \cite{Foster2010}. We thank A.~Baggaley for useful discussions.  This work was supported by EPSRC grants EP/H027777/1 and BH101785.} \\
\appendix 
\section{Error Checking}
\begin{figure}[!htb]
\subfigure[Total Vortex Number]{\includegraphics[width=0.9\linewidth]{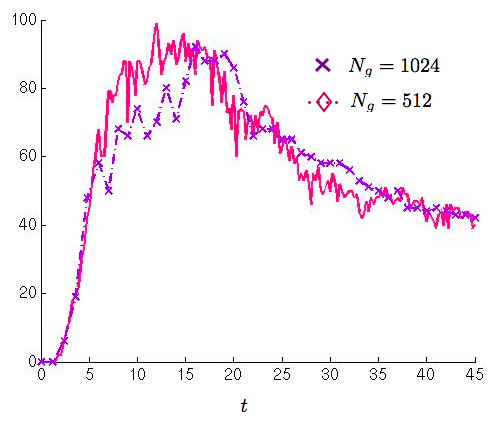}}
\subfigure[Angular Momentum]{\includegraphics[width=0.9\linewidth]{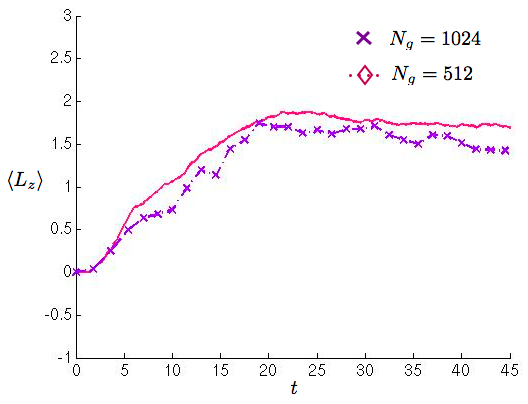}}
\caption{b) Evolution of angular momentum and a) total number of vortices nucleated by a paddle rotating at the center of the condensate ($V_{{\text{I}}}$), with $v=6$, $d=1$.  Simulation gridsizes $N_{g}=512$ (purple) and $N_{g}=1024$ (pink).}
\end{figure}
To check our results are independent of gridsize used, the simulation grid size was doubled from $N_{g}=512$ to $N_{g}=1024$ for a paddle rotating at the condensate center with $v=6$ and $d=1$. In figure 9 a comparison is made of the total vortex number and condensate angular momentum, when the grid size is doubled. The angular momentum is calculated by evaluating Eq~(\ref{angmom}).
The reasonable agreement between rates of vortex production and elimination, as well as evolution of the condensate angular momentum in both runs establishes that the gridsize of $N_{g}=512$ applied in the simulations presented in the body of the paper is adequate. The small variance in results from doubling the gridsize are attributed to the condensate edge selection routine used. A further source of difference is the chaotic nature of vortex dynamics in turbulent systems.  A small amount of numerical noise would be enough to seed a difference in vortex trajectories.  


\bibliography{vortexnucleation}
\end{document}